# A Bright Molecular Core in a Crab Nebula Filament

E.D. Loh and J.A. Baldwin

Department of Physics and Astronomy, Michigan State University, East Lansing, MI 48824-2320 USA
(loh@pa.msu.edu, baldwin@pa.msu.edu)

and

G.J. Ferland

Department of Physics, University of Kentucky, Lexington, KY 40506, USA
(gary@pa.uky.edu)

**Abstract**

In a sub-arcsec near-infrared survey of the Crab Nebula using the new Spartan Infrared Camera, we have found several knots with high surface brightness in the $H_2$ 2.12μm line and a very large $H_2$ 2.12μm to Brγ ratio. The brightest of these knots has an intensity ratio $I(H_2\ 2.12\mu m)/I(Br\gamma) = 18\pm9$, which we show sets a lower limit on the ratio of masses in the molecular and recombination (i.e. ionized) zones $M_{mol} / M_{rec} \geq 0.9$, and a total molecular mass within this single knot $M_{mol} \geq 5\times10^{-5}$ $M_\odot$. We argue that the knot discussed here probably is able to emit so strongly in the 2.12μm line because its physical conditions are better tuned for such emission than is the case in other filaments. It is unclear whether this knot has an unusually large $M_{mol} / M_{rec}$ ratio, or if many other Crab filaments also have similar amounts of molecular gas which is not emitting because the physical conditions are not so well tuned.

*Subject headings:* ISM: molecules – ISM: supernova remnants –supernovae: individual (Crab Nebula)

## 1. Introduction – $H_2$ Emission and the Mass of the Crab Nebula Filaments

The mass of the "recombination" ($H^+$) region in Crab Nebula filaments has been measured from optical spectra starting with O'Dell (1962). It has long been known that this mass is about half the amount expected to have been ejected from a supernova explosion (Hester 2008 and references therein). It has been postulated (Chevalier 1977; Sankrit & Hester 1997) that much of the missing 4–6$M_\odot$ lies in a largely unseen expanding outer shell of the Crab Nebula, but the one apparent detection of this shell only directly shows the presence of about 30 percent of the missing mass (Sollerman et al. 2000).

Here we investigate the possibility that there may be considerably more mass associated with the filaments than is usually appreciated, in the form of hard-to-detect molecular gas. Indeed, the $H_2$ 2.12μm emission line has been detected by Graham et al. (1990; hereafter G90) at two of the three locations that they observed. The G90 measurements were made through a 19 arcsec diameter aperture, which averages over a large and complicated region in the filamentary structure at both locations. Fesen et al (1997) have very roughly estimated the $H_2$ mass, based on ascribing all of the observed extinction to the molecular gas, to be $1.2\pm0.5$ $M_\odot$. But the molecular mass could be very different from that value, in either direction. A complete survey of the molecular content of the whole Crab Nebula is needed to find out.

We report on our detection of a very bright, 2 arcsec diameter knot of $H_2$ emission just outside the area covered by one of G90's aperture positions. We also see additional fainter knots which may have similar properties.

## 2. Observations

### 2.1 Spartan Infrared Camera Images

The observations reported here are from a NIR survey of the Crab Nebula to low surface brightness, at 0.5–1 arcsec angular resolution. We used the new Spartan Infrared Camera (Loh et al. 2004) on the 4.1m SOAR Telescope. Built at Michigan State University, the camera has 4 Rockwell HAWAII2 2K×2K HgCdTe detectors which image the sky in a closely-spaced array. We used its lower-resolution mode which produces a plate scale of 0.068 arcsec pixel$^{-1}$ over a 5×5 arcmin$^2$ field of view. Further observational details will be given by E. D. Loh et al. (2010, in preparation).

Our survey covers the full extent of the Crab Nebula, but the results shown here are for one small region and from only one observing night (2009 December 25 UT). The observations were taken through narrow-band $H_2$, Brγ and continuum filters centered at 2.117μm, 2.162μm and 2.208μm, and with FWHM = 0.030μm, 0.020μm and 0.030μm, respectively. The individual exposures were 180–250 s in length, with chopping to a sky position at 10-minute intervals and with dithering of the positions of the object exposures, following the usual infrared observing procedures. The flux calibration was determined by observing the standard star Persson 9116 (Persson 1998) on the night of 2009 December 2 UT. Comparison to 2MASS stars in the Crab field shows that after conversion from $H_2$ to K magnitudes, our magnitudes are about 0.07 mag too faint.

The region of interest is the filament FK-10 (Fesen & Kirshner 1982), which lies about 70 arcsec SE of the pulsar. In optical emission lines, it is one of the brightest filaments in the Crab, and consequently it has been the subject of many studies. It is one of the two spots on the Crab where previous $H_2$ observations have been made (G90).

Figure 1a shows the sum of 18 individual images covering FK-10 taken through the $H_2$ filter, adding up to 54 min total exposing time. Figures 1b and 1c show the $H_2$ and Brγ images from this same region after subtracting off the continuum. Note that the synchrotron continuum emission that fills much of the field of view in Figure 1a has cleanly subtracted off, showing that our continuum subtraction is accurate.

Several $H_2$ emitting knots are clearly seen in Fig. 1b, with Knot 1 being of special interest because it has much higher surface brightness than the others. A preliminary look over our data for the parts of the Crab Nebula not shown in Fig. 1 found no other knots that, within the passband of our filter, have $H_2$ emission that is nearly as bright as Knot 1.

### 2.2 Optical Passband Data

We used archival HST WFPC2 images (Blair et al. 1997) taken through the F502N, F675N and F547M filters. Fig 1d shows a continuum-subtracted [O III] λ5007 image covering the same region as the infrared images, and Fig. 1e shows the [S II] λ6717+6731 emission lines but in this case without continuum subtraction.

Knot 1 is sufficiently unusual that we took an optical spectrum across it, using the Goodman High Throughput Spectrograph (Clemens et al. 2004) on the SOAR Telescope, with a 1 arcsec slit width and 5Å FWHM wavelength resolution. We calibrated the flux using standard stars from Hamuy et al. (1992). Our long-slit spectrum shows strong emission in the Balmer lines and also in [O III], [O I], [N II] and [S II] lines, which we assume comes from the $H_2$ knot, at heliocentric velocity $v_{helio}$ = +161 km s$^{-1}$ (Figure 1f). There are additional emission features both along the same line-of-sight as the knot and from nearby regions along the slit covering the velocity range – 2800 < $v_{helio}$ < +1600 km s$^{-1}$. All except the very weakest of these features fall at velocities near the peak transmission of our $H_2$ and Brγ filters, so flux measurements for these lines using the IR images should not be strongly affected by the filter passbands. Similarly, the emission features in this part of the Crab all fall at velocities corresponding to the flat tops of the transmission curves for the narrow-band HST filters, so the [O III] λ5007 surface brightness from the continuum-subtracted HST image should be well-calibrated.

**2.3 Observed Properties of the $H_2$ Emission Knot**

Knot 1 is centered at RA (2000) = 05$^h$ 34$^m$ 34.39$^s$, Dec = +21° 59' 40". The total measured $H_2$ 2.12μm flux coming from the part of the knot within the contour having 50 percent of peak surface brightness is 1.3×10$^{-14}$ erg cm$^{-2}$ s$^{-1}$. This falls in an area of 5.4 arcsec$^2$, giving an average surface brightness $S$($H_2$ 2.12μm) = 2.4×10$^{-15}$ erg cm$^{-2}$ s$^{-1}$ arcsec$^{-2}$, with about 10 percent uncertainty. For a distance of 2 kpc (Trimble 1973), the $H_2$ luminosity is $L(H_2)$ = 5.8×10$^{30}$ erg s$^{-1}$.

The $H_2$ emission from the knots is very strong relative to H recombination lines, as is evident from a comparison of the $H_2$ image in Fig. 1b to the Brγ image in Fig. 1c. The small circles drawn in the panels in Fig. 1 mark the locations of several stars, and are intended to make it easier to compare the positions of features visible in the different images. Weak Brγ emission can be traced running from the lower left to the top right, along the presumably photoionized filament that shows up strongly in the [O III] and [S II] lines. There is no $H_2$ emission detected from most of these regions, which is consistent with the intensity ratio $I(H_2\ 2.12μm)/I(Brγ)$ ~ 0.07 that is expected from a PDR and H II region. However, for Knot 1, $I(H_2\ 2.12μm)/I(Brγ)$ = 18 ± 9, with the large uncertainty being due to the very weak Brγ line.

In addition to the $H_2$ 2.12μm line, G90 also measured Brγ from the ionized gas. In both locations where they detected $H_2$, they found $I(H_2\ 2.12μm)/I(Brγ)$ ~ 1. The G90 aperture position for FK-10 is shown on Fig. 1b as a large, dashed circle. It appears that the G90 results average over a mix of regions that produce a very high $H_2$/Brγ ratio and ones that produce the much lower ratio expected from a standard PDR.

Table 1 compares the strengths of the two infrared lines to Hβ, corrected for reddening using $A_V$ =1.6 (Miller 1973). Our optical spectrum covers only part of the area of the knot, so it was necessary to correct for spatial variations in surface brightness. We used the calibrated HST [O III] image to find that the average [O III] surface brightness over the full area covered by Knot 1 is about 0.3 times as large as for the region covered by the extracted spectrum, and have scaled the surface brightness $S$(Hβ) used in Table 1 by this amount so that it can be properly compared to the H$_2$ measurement for the whole knot. This adjustment assumes that the surface brightness in Hβ scales along with that of [O III]. The [O III]/Hβ ratio measured at different locations along our spectrograph slit varies from 5–12, with the value at the position of Knot 1 being 7, near the low end of this range. We estimate that this introduces 30 percent uncertainty into the H$_2$/Hβ ratio.

The HST image shows that there is [O III] emission from the position of Knot 1, but the knot does not especially stand out in the image. However, an unusually bright feature is visible at the position of Knot 1 in the HST [S II] image, and our spectrum shows that [O I] and [S II] are somewhat stronger than elsewhere in the FK10 region.

## 3.0 Analysis

### 3.1 Selection effects and H$_2$ emission

Table 2 shows the $I$(H$_2$ 2.12μm)/$I$(Brγ) line ratio in three environments: the Orion H II region and PDR, a filament near the bright radio galaxy NGC 1275 (Ferland et al. 2008), and the position we observed in this paper. The Crab filament clearly has exceptionally strong H$_2$ emission relative to neighboring H I recombination lines. G90 considered in some detail three possible methods for exciting the H$_2$ emission: UV fluorescence, shocks, and heating by fast electrons.

Here we present a simple argument about the mass of molecular gas needed to produce the observed emission, without needing to know the exact excitation mechanism. Selection effects that determine a line's emissivity also limit the density and temperature where a line can form efficiently. This can set an upper limit to the H$_2$ line emissivity and a lower limit to the molecular mass.

We assume that H$_2$ emission is collisionally excited by thermal particles. Starlight fluorescence excitation, the so-called Solomon process (Osterbrock & Ferland 2006, hereafter AGN3), produces H$_2$ lines that are quite faint relative to H I recombination lines. The Orion Nebula and PDR (Table 2) is an example of the efficiency of such excitation. Photoexcitation cannot produce such strong H$_2$ emission relative to H I.

The filaments near NGC 1275 are excited by penetration of energetic particles from the surrounding hot plasma (Ferland et al. 2009; hereafter F09). G90 proposed a similar explanation for the H$_2$ emission in the Crab. Shock excitation is another obvious candidate as a power source. No matter which of these two is the fundamental energy source, the H$_2$ molecule is excited by collisions with thermal particles. The temperature of the thermal particles is set by the energy source, but the temperature sets the H$_2$ emission properties (F09).

We know little about the properties of the region where the H$_2$ lines form. To make progress we consider a wide range of gas density and kinetic temperature. We have taken the H$_2$ emission predictions from F09 and present them in Figure 2 as the ratio of the volume emissivity of the 2.12 μm line divided by the square of the hydrogen density. This is shown as a function of density and kinetic temperature. The $n^2$ factor converts the volume emissivity into an emission coefficient and is appropriate for a line that is excited by a binary collision. An H I recombination line forms by a collision between a proton and an electron while for H$_2$ the collision is between a hydrogen molecule and another gas constituent. This density-squared dependence holds for densities smaller than the critical density of the upper level. For an H I line this critical density is of order $10^{15}$ cm$^{-3}$ while for the H$_2$ 2.12 μm line the critical density is ~$10^5$ cm$^{-3}$ (F09). We consider densities above the H$_2$ critical density in the following, which makes the line less emissive than we will assume.

The temperature dependence of the H$_2$ 2.12 μm emission coefficient is due to the selection effects discussed in Ferland et al. (2008). At low temperatures the Boltzmann factor for the upper level of the 2.12 μm line, which is ~7000 K above ground, is too small to allow effective excitation. The emissivity goes down for temperatures $T \geq 5000$K because the molecule is collisionally dissociated and the H$_2$ abundance goes down; the H$^0$ formed in this way would be counted as part of the total mass $M_{mol}$ in the region emitting the molecular lines. The upper level of the 2.12 μm transition is collisionally de-excited for densities above the critical density of $10^5$ cm$^{-3}$. The result is a peak emission coefficient $4\pi J(2.12\mu m)/n^2_{mol} \leq 10^{-26}$ erg cm$^3$ s$^{-1}$ at temperatures around 3000K for densities around $10^4$ cm$^{-3}$. We assume this peak emission coefficient in the mass estimates that follow. If the true density and temperature are different from this value, the emission coefficient will be smaller and we will underestimate the mass in the molecular region.

### 3.2 A lower limit to the mass ratio $M_{mol} / M_{rec}$

Here we use the H$_2$ 2.12 μm line to trace $M_{mol} / M_{rec}$, the ratio of the mass in the molecular region to the mass in the region forming recombination lines. We compare the H$_2$ line to the neighboring Br$\gamma$ recombination line to minimize uncertainties due to reddening.

The upper limit of the H$_2$ line relative to Br$\gamma$ is given by

$$\frac{I(2.12\mu m)}{I(Br\gamma)} \leq \frac{n^2_{mol}}{n^2_{rec}} \frac{4\pi J(2.12\mu m)/n^2_{mol}}{4\pi J(Br\gamma)/n^2_{rec}} \frac{V_{mol}}{V_{rec}}$$

$$\approx \frac{n_{mol}}{n_{rec}} \frac{4\pi J(2.12\mu m)/n^2_{mol}}{4\pi J(Br\gamma)/n^2_{rec}} \frac{M_{mol}}{M_{rec}}$$
(1)

Unlike the case for H$_2$, the Br$\gamma$ emissivity is reasonably well-determined. Assuming Case B, the tables in AGN3, and a gas kinetic temperature of $T_{rec} = 15000$ K, typical of ionized gas in a Crab filament, we find a Br$\gamma$ emission coefficient of $4\pi J(Br\gamma)/n^2_{rec} \sim 2.5\times 10^{-27}$ erg cm$^3$ s$^{-1}$. If we assume an emission coefficient of $4\pi J(2.12\mu m)/n^2_{mol} \leq 10^{-26}$ erg cm$^3$ s$^{-1}$ we find the relation

$$\frac{I(2.12\mu m)}{I(Br\gamma)} \leq 4 \frac{n_{mol}}{n_{rec}} \frac{M_{mol}}{M_{rec}}$$
(2)

The density $n_{mol}$ within the molecular region is unknown. The peak H$_2$ emission coefficient occurs at $T \sim 3000$ K. If the temperature is far different from this, the H$_2$ line will be far less emissive than we assumed and a greater molecular mass will be needed. If we assume this temperature, a factor of five below those measured for the H$^+$ region, and constant gas pressure, the density in the molecular region is five times greater than that measured by ionized gas diagnostics. This implies $n_{mol} \sim 10^4$ cm$^{-3}$, near the peak H$_2$ emission coefficient. A density either much lower or higher than we assume would decrease the emission coefficient and *increase* the mass ratio. Assuming that the ratio of densities is five we find

$$\frac{M_{mol}}{M_{rec}} \geq 0.05 \frac{I(2.12\mu m)}{I(Br\gamma)} \tag{3}$$

The observed ratio of 18 corresponds to a mass ratio of $M_{mol} / M_{rec} \geq 0.9$.

Because the Brγ emission from Knot 1 is close to our detection limit, we have checked our result by also measuring the H$_2$ 2.12μm/Hβ ratio for Knot 1 (Table 1). Using Hβ in place of Brγ in the procedure described above, this ratio implies $M_{mol} / M_{rec} \geq 2$, but now with considerable sensitivity to reddening and to any relative errors between the optical and NIR flux calibrations. The Brγ/Hβ intensity ratio given in Table 1 is in fact about twice the Case B value, suggesting that Brγ is actually weaker than the value adopted here. Although a H$_2$/Brγ intensity ratio two times smaller than 18 is within the error bars, the Hβ result suggests that it is more likely that any error in the intensity and hence in the lower limit on the mass ratio is in the upwards direction. In the following discussion, we use $M_{mol} / M_{rec} \geq 0.9$. The corresponding mass of H$_2$ in Knot 1 is
$M_{mol} = [m_H n_{mol} L(2.12\mu m)]/[4\pi J(2.12\mu m)] \geq 5 \times 10^{-5}$ M$_\odot$, where $m_H$ is the mass of a hydrogen atom. This mass limit is insensitive to $n_{mol}$ for $n_{mol} \geq 10^4$ cm$^{-3}$.

**4.0 Discussion and Conclusions**

It is important to find out why Knot 1 (and to a lesser extent the other H$_2$ knots visible in Figure 1b) are so bright in H$_2$ emission. It may be that Knot 1 just has an unusually large amount of molecular gas. Indeed, its optical spectrum shows anomalously strong [S II] and [O I] lines, and it is part of the FK-10 complex which is one of the optically-brightest knots in the Crab.

Or perhaps the H$_2$ emission from Knot 1 is telling us that there are significant amounts of undetected molecular gas in other Crab filaments. The H$_2$ emissivity of Knot 1 must be somewhere around the peak value shown in Figure 2. A maximally-emitting cube 2.3 arcsec on a side at the distance of the Crab would emit $3 \times 10^{32}$ erg s$^{-1}$ in the H$_2$ line, so the observed H$_2$ luminosity $L(2.12\mu m) = 6 \times 10^{30}$ erg s$^{-1}$ corresponds to such a cube emitting at 2 percent of the maximum rate, or alternatively to a maximally-emitting sheet fifty times thinner along the line of sight than its projection on the sky. The emissivity is a steep function of temperature and is sensitive to density above the 2.12 μm line's critical density. If other knots were not precisely tuned to the right maximally emissive parameters, their H$_2$ emissivity would go down. In that case, there could be an as-yet unknown mass of cold gas in the cores of other filaments, but which does not produce strong 2.12μm emission. We also stress that even for Knot 1 our $M_{mol}$ estimate is a lower limit, and there could easily be much more molecular gas present.

The observed dust content of the whole Crab is the strongest present constraint on the total amount of molecular gas. Our Cloudy simulations show that normal Galactic dust associated with molecular gas like that in Knot 1 should emit at about 200μm. The observed thermal emission peaking at around 80 μm (Green et al. 2004; Temim et 2006) must come from about $10^{-2}$ $M_\odot$ of warmer dust in the recombination region. For a typical dust/gas ratio of about 100, the limit of $< 0.1 M_\odot$ of cold dust set by the 170μm ISO data (Green et al 2004) does not place an interesting limit on the amount of associated molecular gas. Knot 1 is on the far side of the Crab's expanding shell, as is shown by its radial velocity. If cold dust is common we should see it in absorption from other filaments on the near side. Patchy continuum absorption due to dust is in fact seen (Fesen & Blair 1990; Hester et al. 1990), but the estimated mass of such dust is very small ($M_{dust} \leq 10^{-2} M_\odot$).

We will explore these issues further in future papers. Our next steps will be to take a near-IR spectrum across Knot 1 to measure the $H_2$ temperature, and to more thoroughly map the optical emission lines in this region. Our near-IR imaging survey is intended to complement the existing Spitzer images and spectra (Temim et al 2006) which include much of the Crab Nebula. The Spitzer InfraRedSpectrograph spectra cover a number of pure-rotational $H_2$ lines which would constrain the amount of cooler $H_2$ gas in other filaments. We are also planning to add new grids of optical spectra at the Spitzer spectroscopic slit locations. It is important to bring high angular resolution optical and near-IR measurements together with the existing Spitzer data base, to provide a broad set of observational constraints on the nature of the molecular cores in the Crab filaments.

*Acknowledgements.* We are grateful to Roger Chevalier, James Graham, Rob Fesen, Bob O'Dell and an anonymous referee for valuable comments. EL, JAB and GJF acknowledge support from NASA ADP grant NNX10AC93G. GJF acknowledges support from NSF 0607028 and 0908877 and by NASA through 07-ATFP07-0124.

| Table 1. Reddening Corrected Intensity Ratios for Knot 1 | |
|---|---|
| Line | $100 \times I/I(H\beta)$ |
| $H_2$ 2.12μm | 125±40 |
| H I Brγ 2.17 μm | 7±4 |
| $S(H\beta) = 2.31\times10^{-15}$ erg cm$^{-2}$ s$^{-1}$ arcsec$^{-2}$ | |

| Table 2. Intensities of $H_2$ relative to H I in three objects | |
|---|---|
| Object | $I(H_2\ 2.12\mu m)/I(Br\gamma)$ |
| Orion | 0.07 |
| NGC 1275 | 8 |
| Crab filament | 18 |

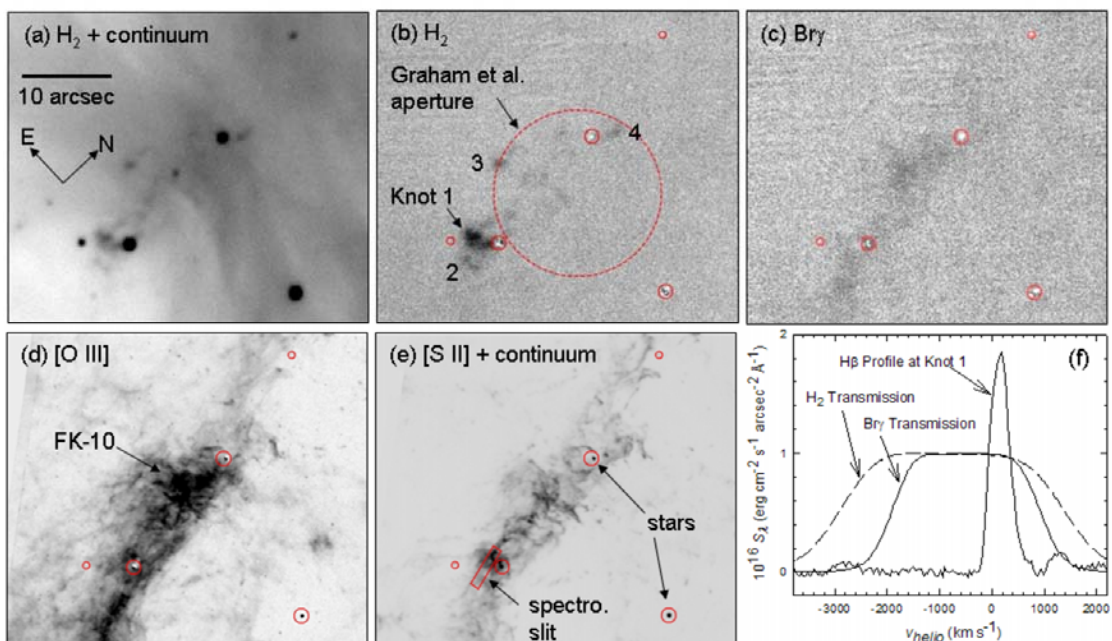

**Figure 1.** Panels (a)-(e) show the same region including the $H_2$-emitting knots discussed here. The small circles mark the positions of several stars that can be seen in panel (a), for use as reference points. Panel (f) shows the Hβ region of the optical spectrum.

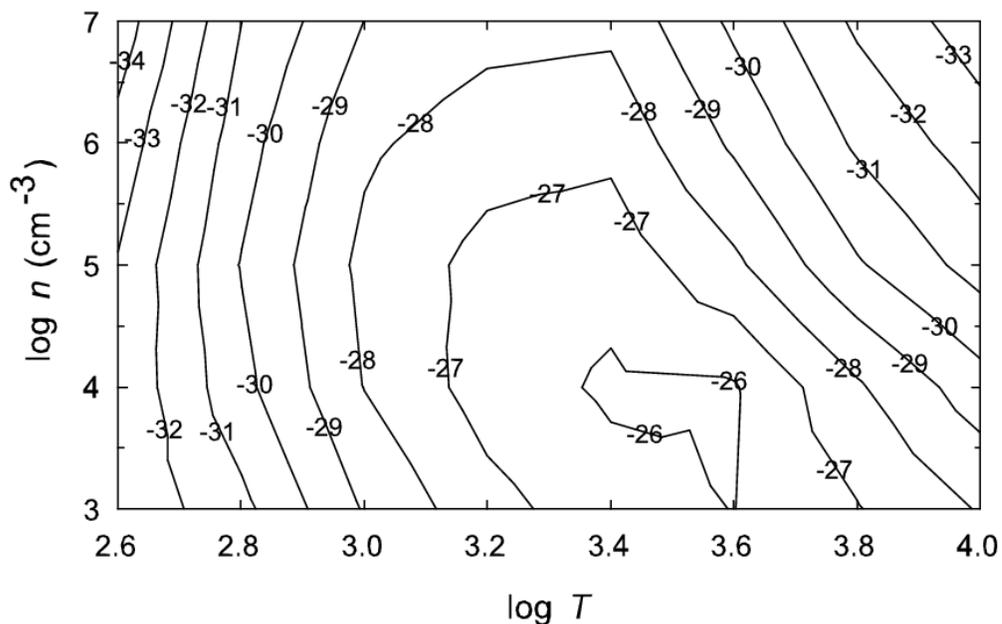

**Figure 2.** Contours of $\log_{10}$ of the emission coefficient $4\pi\,J(2.12\mu m)/n^2$ (erg cm$^3$ s$^{-1}$), where $n$ is the density of H baryons in all forms and $T$ is the gas kinetic temperature. This allows an $H_2$ 2.12μm intensity to be converted into a *total* mass in the $H_2$-emitting volume.